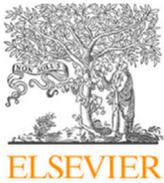

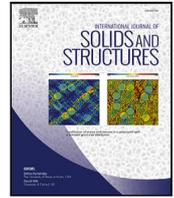

# Adhesive contact mechanics of viscoelastic materials

C. Mandriota [a], N. Menga [a,*], G. Carbone [a,b,c]

[a] *Department of Mechanics, Mathematics and Management, Politecnico di Bari, Via Orabona, 4, 70125, Bari, Italy*
[b] *Imperial College London, Department of Mechanical Engineering, Exhibition Road, London SW7 2AZ, United Kingdom*
[c] *CNR - Institute for Photonics and Nanotechnologies U.O.S. Bari, Physics Department "M. Merlin", via Amendola 173, 70126 Bari, Italy*

## A R T I C L E   I N F O

## A B S T R A C T

In this study, we propose a theory of rough adhesive contact of viscoelastic materials in steady-state sliding. By exploiting a boundary formulation based on Green's function approach, the unknown contact domain is calculated by enforcing the local energy balance at the contact edge, thus considering also the non-conservative work of internal stresses which is directly related to the odd part of the Green's function. Theoretical predictions indicate that viscoelasticity may enhance the adhesive performance depending on the sliding velocity, thus leading to larger contact area and pull-off force compared to the equivalent adhesive elastic case The interplay between viscoelasticity and adhesion also affects the overall friction. Indeed, at low velocity, friction is strongly enhanced compared to the adhesiveless viscoelastic case, mainly due to the small-scale viscoelastic hysteresis induced by the adhesive neck close to the contact edge At higher velocity, the effect of viscoelastic hysteresis occurring at larger scales (bulk material) leads to even higher friction. Under these conditions, in the presence of adhesion, the small-scale and large-scale viscoelastic contributions to friction cannot be separated. Finally, in contrast with usual predictions for crack propagation/healing in infinite systems, we found a non-monotonic trend of the energy release rates at the trailing and leading contact edges, which is consistent with the finiteness of the contact length. All the presented results are strongly supported by existing experimental evidences.

## 1. Introduction

Understanding the effect of adhesion in sliding contacts of polymeric rubber-like materials is of crucial importance in a large number of engineering and industrial applications, as in Micro-Electro-Mechanical systems (MEMs) (Bhushan, 2007; Kim et al., 2007; Zeng et al., 2018), medical applications (Vilhena and Ramalho, 2016; Van den Dobbelsteen et al., 2007), seals (Tiwari et al., 2017; Gawliński, 2007), biological and bio-inspired systems (Chung and Chaudhury, 2005; Zhou et al., 2013; Tang et al., 2008; Menga et al., 2018b, 2020; Ceglie et al., 2022), protective coatings (Martinez-Martinez et al., 2012; Zhang and Archer, 2007), tire–road frictional behavior (Sharp et al., 2016; Persson, 2000), windscreen wipers (Bódai and Goda, 2014), touchscreens (Ayyildiz et al., 2018), and solid state batteries (Zhang et al., 2023). Controlling adhesion is fundamental, as the adhesive effect might be desirable, e.g. adhesion might increase the sliding friction in the tire–road contact thus improving the braking and the handling performances (Sharp et al., 2016), or deleterious, e.g. in MEMs and micro-grippers, where avoiding the permanent adhesion is a crucial issue for moving components and to release the objects (Zhang et al., 2009; Mengüç et al., 2012; Zeng et al., 2018).

Viscoelasticity is another major feature affecting the contact behavior of rubber-like materials. Existing analytical and numerical approaches (Hunter, 1961; Harrass et al., 2010; Persson, 2001, 2010; Carbone and Putignano, 2013; Menga et al., 2014, 2016b, 2018a; Putignano et al., 2019; Afferrante et al., 2019) are only able to describe the adhesiveless viscoelastic contact behavior in the presence of relative motion between the solids, also considering normal–tangential coupling in the deformation field due to interfacial friction and finite thickness (Menga, 2019; Menga et al., 2021; Müller et al., 2023). Indeed, under these conditions, viscoelastic relaxation strongly affects the key contact quantities, such as the contact size, which are found to depend on the relative sliding velocity. Moreover, since viscoelasticity also entails a certain amount of energy dissipation within the bulk material under the effect of cyclic deformations, in these class of contacts, a friction force is observed opposing the relative interfacial sliding, which is usually referred to as viscoelastic friction.

Although several experimental results (Barquins et al., 1978; Charmet and Barquins, 1996; Hoyer et al., 2022) clearly indicate that interfacial adhesion and viscoelasticity may interact in controlling the properties of mating interfaces in many tribological systems, a comprehensive theory of adhesive viscoelastic

---

sliding contacts is still lacking. Indeed, the effect of adhesion on the behavior of contacting solids has been mostly investigated only for purely elastic solids, by relying on both theoretical (Johnson et al., 1971; Maugis, 2000; Menga et al., 2018c, 2019) and numerical (Carbone and Mangialardi, 2008; Carbone et al., 2009b, 2015; Menga et al., 2016a) approaches based on the Griffith's energy balance. Very little has been done in the case of adhesive viscoelastic contacts, with some contributions primarily focusing on the case of normal indentation and exploiting adhesive models based on surface potentials (Violano and Afferrante, 2022; Pérez-Ràfols et al., 2023).

In the case of viscoelastic materials, the standard energy balance equation $G = \Delta\gamma$ between the energy release rate $G$ and the Duprè work of adhesion $\Delta\gamma$, no longer holds true, as additional non conservative energy terms come into play due to viscoelastic hysteresis (Persson and Brener, 2005; Schapery, 1975; Barber et al., 1989; Schapery, 1989; Greenwood, 2004; Baney and Hui, 1999; Carbone and Persson, 2005a,b). Moreover, in the presence of adhesion, besides the usual large-scale energy dissipation occurring in the bulk viscoelastic material, local dissipation also takes place very close to the contact edges (i.e., at the opening or closing crack tips). This latter phenomenon, also known as small-scale viscoelasticity, is usually regarded as a primary cause of enhanced adhesive hysteresis, as observed in Fuller and Roberts (1981) and Kendall (1975). Since in these conditions the real value of $G$ at the contact edges is unknown, a very few studies exist focusing on such case. Most of them rely on scale separation (Carbone and Mangialardi, 2004; She et al., 1998; Krijt et al., 2014; Zhang et al., 2015), thus assuming purely elastic conditions in the bulk material, and local crack propagation criterion at the contact edges (Greenwood, 2004; Carbone and Persson, 2005a,b). However, although very pioneering, these studies are limited to the local viscoelastic regime (i.e., small-scale viscoelasticity).

On the other hand, several experimental investigations have reported strongly enhanced adhesive properties in rolling contacts against rubber (i.e., viscoelastic) substrates (Barquins et al., 1978; Charmet and Barquins, 1996; Hoyer et al., 2022). Larger pull-off force and contact size are reported in rolling conditions, compared to the static case, thus confirming that the effective adhesive behavior depends on the interaction between small-scale viscoelasticity, interfacial adhesion, and bulk viscoelasticity (Roberts, 1979). Moreover, also friction is affected by similar mechanisms, as shown in Refs. Grosch (1963) and Roberts (1979) More specifically, Grosch's experiments clearly show that the interplay between viscoelasticity and adhesion leads to a friction increase, which cannot be explained by simply summing-up the contributions of adhesion hysteresis and bulk viscoelastic losses (Persson, 1998; Scaraggi and Persson, 2015). Such an experimental evidence has not yet found any physically explanation.

We propose a novel theory to study the adhesive contact of viscoelastic materials in steady-state sliding or rolling contact with a rigid substrate. The theory provides the closure equations needed to determine the unknown contact domain, which is expressed in terms of a local energy balance, thus generalizing the Griffith's criterion to the case of viscoelastic media. The very first results of the proposed theory have been shown by Carbone et al. in a short letter (Carbone et al., 2022), where the authors report a very good agreement between theoretical prediction and experimental evidences. In this paper, the authors present their theory in much deeper detail, thus providing an exhaustive treatment of viscoelastic adhesive contacts, and presenting a specific analysis of the influence of the physical parameters affecting the tribological behavior of the contact. The theory covers a very wide range of sliding velocities, thus providing insights and enlightening most of the available experimental observations in the field.

## 2. Formulation

We consider a linear homogeneous viscoelastic slab sliding past a rough surface in the presence of interfacial adhesion. We assume steady state motion at constant velocity $\mathbf{v}$, and displacement controlled conditions by assigning the contact penetration $\Delta$. We neglect the presence of shear stresses at the interface. According to Menga et al. (2014) and Carbone and Putignano (2013), using a reference frame co-moving with the indenter, the contact normal stress $\sigma(\mathbf{x})$ and the normal interfacial displacement $u(\mathbf{x})$ (see also Carbone and Mangialardi, 2008 and Menga et al., 2016b for the definition of interfacial displacement) are related each other through the integral relation

$$u(\mathbf{x}) = \int d^2 x_1 \mathcal{G}(\mathbf{x} - \mathbf{x}_1, \mathbf{v})\sigma(\mathbf{x}_1) \tag{1}$$

where $\mathbf{x}$ and $\mathbf{x}_1$ are the in-plane position vectors. Under the assumption of infinitely short-range adhesive interactions, out of the yet unknown contact domain $\Omega$ of size $|\Omega| = A$ the normal stresses vanish, so that the integral in Eq. (1) can be extended to $\Omega$. The kernel $\mathcal{G}(\mathbf{x}, \mathbf{v})$ is the viscoelastic Green function (parametrically dependent on the sliding velocity $\mathbf{v}$), which has been determined for several geometric configurations (i.e., displacement or stress boundary conditions) both for periodic and non-periodic contacts, as a function of the slab thickness (see Appendix A). Now observe that for a viscoelastic material $\mathcal{G}(\mathbf{x}, \mathbf{v})$ is an asymmetric function of $\mathbf{x}$ and can be decomposed into an even (symmetric) $\mathcal{G}^{\mathrm{E}}(\mathbf{x}, \mathbf{v})$ part, and odd (anti-symmetric) $\mathcal{G}^{\mathrm{O}}(\mathbf{x}, \mathbf{v})$ one

$$\mathcal{G}(\mathbf{x}, \mathbf{v}) = \mathcal{G}^{\mathrm{E}}(\mathbf{x}, \mathbf{v}) + \mathcal{G}^{\mathrm{O}}(\mathbf{x}, \mathbf{v}) \tag{2}$$

with $\mathcal{G}^{\mathrm{E}}(\mathbf{x}, \mathbf{v}) = \frac{1}{2}[\mathcal{G}(\mathbf{x}, \mathbf{v}) + \mathcal{G}(-\mathbf{x}, \mathbf{v})]$ and $\mathcal{G}^{\mathrm{O}}(\mathbf{x}, \mathbf{v}) = \frac{1}{2}[\mathcal{G}(\mathbf{x}, \mathbf{v}) - \mathcal{G}(-\mathbf{x}, \mathbf{v})]$. Notably, for a given penetration $\Delta$ of the indenter, the displacement field $u(\mathbf{x})$ is prescribed at points located in the contact domain $\Omega$ where the deformed slab shape must match the rigid profile. Therefore, Eq, (1) can be inverted to calculate the stress field $\sigma(\mathbf{x})$ in the contact domain $\Omega$, being $\sigma(\mathbf{x}) = 0$ out of the contact. The problem at hand belongs indeed to the class of mixed value problems. In fact, out of the contact area the unknown is no longer the normal stress distribution but the displacement field which can simply be calculated directly from Eq. (1) for $\mathbf{x} \notin \Omega$. However, the contact area $A = |\Omega|$ is not yet known, hence an additional equation (the closure condition) needs to be found to completely solve the problem. In the case of adhesiveless contacts, since the local contact pressure can only takes positive values or vanish, it is enough to enforce the condition that the stress $\sigma(\mathbf{x}) = 0$ at the boundary $\partial\Omega$ of the contact domain. In presence of adhesion, instead, the local pressure can also take negative values, i.e. the stress $\sigma(\mathbf{x})$ may change sign over the contact area. Under such conditions, the closure equation can be easily found only for the case of elastic contacts, by requiring that (at fixed penetration) the total energy $\mathcal{F} = U - \Delta\gamma A$ (i.e. the sum of the elastic energy $U$ stored into the system and the adhesion energy $-\Delta\gamma A$, with $\Delta\gamma$ being the Duprè work of adhesion) is stationary at equilibrium (Carbone and Mangialardi, 2008; Maugis, 2000; Menga et al., 2016a, 2018c, 2019). However, additional non-conservative energy contributions exist in viscoelastic materials, which prevent such an approach from being employed. In this case, we need to write a local energy balance, which requires that, at fixed penetration $\Delta$, the work $\delta L$ done by internal viscoelastic stresses equates the work done by external adhesion forces when the contact area is subjected to a small quasi-static perturbation $\delta A$, i.e.

$$\delta L = \Delta\gamma \delta A \tag{3}$$

Due to the hysteretic behavior of the viscoelastic material, $\delta L$ consists of two terms: the change of elastic energy stored in the viscoelastic material, and the work done by non-conservative internal stresses.

In order to provide a physical framework to determine these two contributions, we firstly consider the simpler case represented in Fig. 1, where the free boundary of a constrained deformable linear solid, is loaded with two forces $F_1$ and $F_2$. The resulting displacement $u_1$ and $u_2$ are given by $F_i = K_{ij} u_j$ (we use the Einstein notation for the repeated index), where $K_{ij}$ is the generic response matrix, which we assume to be asymmetric. Notably, $K_{ij} = K_{ij}^{\mathrm{E}} + K_{ij}^{\mathrm{O}}$ with $K_{ij}^{\mathrm{E}} = \frac{1}{2}(K_{ij} + K_{ji})$ and $K_{ij}^O = \frac{1}{2}(K_{ij} - K_{ij}^T)$ being the symmetric and anti-symmetric parts of $K_{ij}$ respectively.

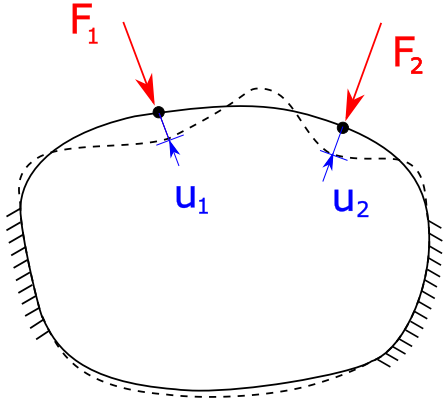

**Fig. 1.** A constrained deformable solid, whose response matrix is non-symmetric, loaded with two surface forces. Blue arrows refer to displacements ($u_1$ and $u_2$), red arrows refer to forces ($F_1$ and $F_2$).

In this case, the work done by $F_1$ and $F_2$ can be written as $\delta L = F_i \delta u_i = K_{ij}^E u_j \delta u_i + K_{ij}^O u_j \delta u_i$, with $\delta u_i$ being the point displacements infinitesimal changes. With this regard, we observe that the symmetry of $K_{ij}^E$ implies that $K_{ij}^E u_j \delta u_i = K_{ij}^E u_i \delta u_j$, and $K_{ij}^E u_j \delta u_i = \delta \left(\frac{1}{2} K_{ij}^E u_i u_j\right)$. Thus, $K_{ij}^E u_j \delta u_i = \delta U$ is a conservative term, i.e. the change of the elastic energy $U = \frac{1}{2} K_{ij}^E u_i u_j$. Similarly, since $K_{ij}^O$ is anti-symmetric, we have that $K_{ij}^O u_j \delta u_i = -K_{ij}^O u_i \delta u_j$, and $K_{ij}^O u_j \delta u_i = \frac{1}{2} K_{ij}^O \left(u_j \delta u_i - u_i \delta u_j\right)$ which cannot be derived by a potential energy. Therefore, the quantity

$$\delta L_P = K_{ij}^O u_j \delta u_i \tag{4}$$

is a non-conservative path-dependent contribution to $\delta L$. Moreover, since $K_{ij}^O u_j u_i = -K_{ij}^O u_i u_j = 0$, the potential energy $U$ can also be rewritten as

$$U = \frac{1}{2}\left(K_{ij}^E + K_{ij}^O\right) u_j u_i = \frac{1}{2} F_i u_i \tag{5}$$

and

$$\delta L = \delta U + \delta L_P \tag{6}$$

The same arguments can be extended to the aforementioned continuum case, i.e. the sliding or rolling contact between a rigid rough indenter and a linear viscoelastic slab (see Appendix B). Now, without any loss of generality, let us consider the case of a generic contact area increase from $A$ to $A + \delta A$ (i.e., $\delta A$ is a positive perturbation). Since we assume infinitely short range adhesive forces (i.e., JKR limit for large values of the Tabor parameter (Johnson et al., 1971)), no interactions occur out of the contact area. Under these conditions, the contact area variation represents a mode I crack closure, and the work $\delta L$ only results from the interfacial stress acting in the small region $\delta A$. Then, within $\delta A$, the $u(\mathbf{x})$ undergoes a quasi-static change $\Delta u(\mathbf{x})$ from the initial shape (out of contact) $u_-(\mathbf{x}) = u(\mathbf{x}, A)$, to final one (in contact) $u_+(\mathbf{x}) = u(\mathbf{x}, A + \delta A)$. This zipping process can be described by introducing a dimensionless parameter $\eta$ that slowly increases from zero to one; therefore, within $\delta A$, we have $u(\mathbf{x}, \eta) = \eta \Delta u(\mathbf{x}) + u_-(\mathbf{x})$, where $\Delta u(\mathbf{x}) = u_+(\mathbf{x}) - u_-(\mathbf{x})$, and $\eta \in [0, 1]$. Linearity entails a similar trend for the stress in $\delta A$, thus $\sigma(\mathbf{x}, \eta) = \eta \sigma_+(\mathbf{x})$, where $\sigma_+(\mathbf{x}) = \sigma(\mathbf{x}, A + \delta A)$. Observing that $\delta u(\mathbf{x}, \eta) = \Delta u(\mathbf{x}) \delta \eta$, the internal work $\delta L = \int d^2 x \sigma(\mathbf{x}) \delta u(\mathbf{x})$ can be rewritten as

$$\delta L = \int_{\delta A} d^2 x \int_0^1 \frac{\partial u}{\partial \eta}(\mathbf{x}, \eta) \sigma(\mathbf{x}, \eta) \delta \eta = \frac{1}{2} \int_{\delta A} d^2 x \Delta u(\mathbf{x}) \sigma_+(\mathbf{x}) \tag{7}$$

Since $\sigma_-(\mathbf{x}) = \sigma(\mathbf{x}, A) = 0$ for $\mathbf{x} \in \delta A$, $\sigma_+(\mathbf{x}) = \sigma_-(\mathbf{x}) = 0$ for $\mathbf{x} \notin A + \delta A$, and $\Delta u(\mathbf{x}) = 0$ for $\mathbf{x} \in A$, the integral in Eq. (7) can be extended to the entire nominal contact area, thus yielding

$$\delta L = \frac{1}{2} \int d^2 x [u_+(\mathbf{x}) - u_-(\mathbf{x})][\sigma_+(\mathbf{x}) + \sigma_-(\mathbf{x})] \tag{8}$$

Moreover, recalling that, according to the aforementioned arguments, the elastic energy is given by $U = \frac{1}{2} \int d^2 x \sigma(\mathbf{x}) u(\mathbf{x})$, we have

$$\delta U = \frac{1}{2} \int d^2 x [u_+(\mathbf{x}) \sigma_+(\mathbf{x}) - u_-(\mathbf{x}) \sigma_-(\mathbf{x})] \tag{9}$$

which only depends on the symmetric part $\mathcal{G}^E(\mathbf{x}, \mathbf{v})$ of the Green's function (see also Ref. Carbone et al., 2022).

Using Eqs. (8) and (9) in Eq. (6), the non-conservative term $\delta L_P$ can be then calculated as

$$\delta L_P = \frac{1}{2} \int d^2 x \left[u_+(\mathbf{x}) \sigma_-(\mathbf{x}) - u_-(\mathbf{x}) \sigma_+(\mathbf{x})\right] \tag{10}$$

which can be shown to strictly depend on the antisymmetric part $\mathcal{G}^O(\mathbf{x}, \mathbf{v})$ of the Green's function (see also Ref. Carbone et al., 2022). Eq. (10) shows that $\delta L_P$ is a non-conservative term, which vanishes in the case of purely elastic material, i.e., when $\mathcal{G}^O(\mathbf{x}, \mathbf{v}) = 0$ (Carbone et al., 2022; Carbone and Mangialardi, 2008; Menga et al., 2016a). Moreover, it is worth noting that this term does not represent the amount of energy dissipated within the bulk viscoelastic material. Indeed, Eq. (10) shows that the sign of $\delta L_P$ can be either positive or negative, depending on the specific conditions. The reader is referred to Appendix D for further insides on conservative and non-conservative mechanical systems, where some special cases yielding $\delta L_P = 0$ are described.

Combining Eq. (6) with Eq. (3), the final expression for the energy balance at the boundary of the contact area (i.e., the closure equation for the unknown contact area) reads

$$\left.\frac{\partial U(\mathbf{v})}{\partial A}\right|_\Delta + \left.\frac{\delta L_P(\mathbf{v})}{\delta A}\right|_\Delta = \Delta \gamma \tag{11}$$

which can be numerically computed based on contact stress and displacement fields using Eqs. (9), (10). Also, recalling the usual definition of the energy release rate $G = (\partial U / \partial A)_\Delta$ at fixed displacement of the indenter, we get

$$G(\mathbf{v}) = \Delta \gamma - \left.\frac{\delta L_P(\mathbf{v})}{\delta A}\right|_\Delta \tag{12}$$

Notably, depending on the sign of the term $\delta L_P$, the energy release rate can either increase or decrease compared to $\Delta \gamma$, and can even vanish or become negative. When this happens adhesion is switched off, being totally masked by viscoelasticity. Notably, such a mechanism might (at least) partially explain the recently reported ability to control adhesion strength by means of mechanical micro-vibrations (Shui et al., 2020).

Also note that in this treatment we do not take into account the influence of the highly non linear phenomena occurring very close to the tip of the contact edges which may modify the effective energy of adhesion (Yoshizawa and Israelachvili, 1993; Tiwari et al., 2017; Yoshizawa et al., 1993; Maeda et al., 2002; Chaudhury, 1996; Chernyak and Leonov, 1986). These effects can be included in the present theory by replacing $\Delta \gamma$ with an apparent energy of adhesion $G_0$ measured at very low sliding velocity.

## 3. The case of a sinusoidal rigid indenter

In this section we discuss the 1D+1D adhesive periodic contact of a viscoelastic half-plane sliding at constant velocity $v$ against a sinusoidal rigid indenter of wavelength $\lambda$, amplitude $\Lambda$ and wave vector $k = 2\pi/\lambda$. All the geometrical parameters needed to define the problem are shown in Fig. 2. Specifically, we define the contact penetration as $\Delta$. The contact domain is defined by the quantities $l_1$ and $l_2$, which represent the distance of the two contact edges (respectively trailing edge and leading edge) from summit of the sinusoidal indenter. The semi-width of the contact length is $a = (l_1 + l_2)/2$ and the eccentricity is $e = (l_2 - l_1)/2$. In this 1D+1D case (Menga et al., 2014), the Green's function is

$$\mathcal{G}(x, v) = J(0) \frac{2(1 - \nu^2)}{\pi} \log\left|2 \sin \frac{kx}{2}\right|$$

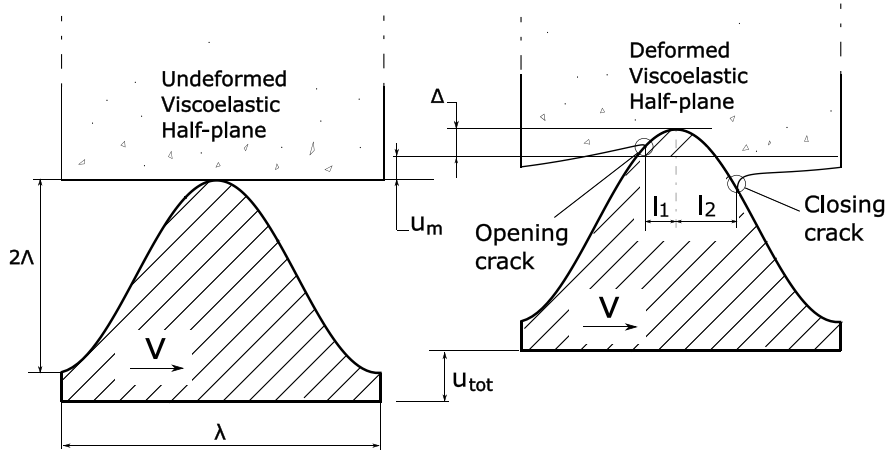

**Fig. 2.** The schematic of the sliding contact between a viscoelastic solid and a rigid wavy indenter. Geometric parameters are also shown.

$$+ \frac{2\left(1-\nu^2\right)}{\pi} \int_{0^+}^{+\infty} dt \log \left|2 \sin \frac{k(x+vt)}{2}\right| \dot{J}(t) \tag{13}$$

The linear viscoelastic solid is modeled with one relaxation time $\tau$, and the creep function is

$$J(t) = \mathcal{H}(t) \left[\frac{1}{E_0} - \left(\frac{1}{E_0} - \frac{1}{E_\infty}\right) \exp\left(-\frac{t}{\tau}\right)\right] \tag{14}$$

where $E_\infty$ and $E_0$ are, respectively, the high and low frequency viscoelastic modulus, and $\mathcal{H}(t)$ is the Heaviside unit-step function.

Within the contact area, the interfacial displacement must match the indenter shape; therefore, $u(x) = \Lambda \cos(kx) - \Lambda + \Delta$ and the contact stress distribution can be found, at given $\Delta$, by solving the equation

$$\int_{-l_1}^{l_2} dx_1 \, \mathcal{G}\left(x - x_1, v\right) \sigma(x_1) = \Lambda \cos(kx) - \Lambda + \Delta \tag{15}$$

In order to calculated the unknown contact parameters $l_1$ and $l_2$ we need to enforce the energy balance Eq. (11) at each edge of the contact, i.e.

$$\left.\frac{\partial U(v)}{\partial l_1}\right|_{\Delta, l_2} + \left.\frac{\delta L_{P_1}(v)}{\delta l_1}\right|_{\Delta, l_2} = \Delta\gamma \tag{16}$$

$$\left.\frac{\partial U(v)}{\partial l_2}\right|_{\Delta, l_1} + \left.\frac{\delta L_{P_2}(v)}{\delta l_2}\right|_{\Delta, l_1} = \Delta\gamma \tag{17}$$

with $\delta l_1$ and $\delta l_2$ being, respectively, the infinitesimal independent variations of the contact area at the trailing and leading edges. The displacement field and the contact stresses are numerically calculated by relying on the numerical procedure addressed in Carbone and Mangialardi (2008).

Once the contact problem is solved [notably, the contact pressure is $p(x) = -\sigma(x)$], we can calculate the remote average pressure

$$p_\infty = \frac{1}{\lambda} \int_{-l_1}^{l_2} p(x)dx, \tag{18}$$

as well as the friction coefficient

$$\mu = -\frac{1}{\lambda p_\infty} \int_{-l_1}^{l_2} p(x) u'(x) dx, \tag{19}$$

and the strain energy release rates at the opening and closing cracks (trailing and leading edges) respectively

$$G_1(v) = \Delta\gamma - \left.\frac{\delta L_{P_1}(v)}{\delta l_1}\right|_{\Delta, l_2} \tag{20}$$

$$G_2(v) = \Delta\gamma - \left.\frac{\delta L_{P_2}(v)}{\delta l_2}\right|_{\Delta, l_1} \tag{21}$$

## 4. Results

Results are shown in terms of the following dimensionless parameters: $\tilde{x} = kx$; $\tilde{a} = ka$; $\tilde{e} = ke$; $\tilde{p} = 2\left(1-\nu^2\right)p/\left(E_0 k\Lambda\right)$; $\tilde{\gamma} = \left(1-\nu^2\right)k\Delta\gamma/(\pi E_0\Lambda)$; $\zeta = kv\tau$; $\tilde{u} = u/\Lambda$; $\tilde{\Lambda} = k\Lambda$; $\tilde{\Delta} = \Delta/\Lambda$; $\beta = E_\infty/E_0$. Also we define the dimensionless elastic energy as $\tilde{U} = 2(1-\nu^2)U/(E_0\Lambda^2)$ and the dimensionless non-conservative work of internal stresses as $\delta\tilde{L}_P = 2(1-\nu^2)\delta L_P/(E_0\Lambda^2)$. Eqs. (16), (17), once written in dimensionless form, show that adhesion is governed by the parameter $\Gamma = \tilde{\gamma}/\tilde{\Lambda}^2$. More specifically, the contact solution is uniquely determined by the parameters $\Gamma$, $\zeta$, $\tilde{\Delta}$ or, analogously, by $\Gamma$, $\zeta$, $\tilde{p}_\infty$. Moreover, regarding the assumption of JKR-like infinitely short-range adhesive interactions, we observe that the Tabor parameter in the stiffer case of $E = E_\infty$ can be rewritten according to our dimensionless quantities as $\mu_T = [(\pi\Gamma/\beta)^2\tilde{\Lambda}^3/(kZ_0)^3]^{1/3}$ and still gives $\mu_T \gg 1$ even for low values of the reduced adhesion energy $\Gamma\ (\approx 0.001)$ and high values of $\beta\ (\gg 10)$ provided that the periodic profile's wavelength $\lambda \gtrsim 10\ \mu m$, which is a reasonable value for our case of interest (also, we set $\tilde{\Lambda} \approx 1$ to enforce linear elasticity and $Z_0 \approx 1$ nm).

Fig. 3 shows the dimensionless semi-width of the contact $\tilde{a}$ [see Figs. 3(a) and (b)] and the dimensionless eccentricity $\tilde{e}$ [see Figs. 3(c) and (d)] as a function of the dimensionless sliding velocity $\zeta = kv\tau$, either at fixed penetration $\tilde{\Delta}$ or load $\tilde{p}_\infty$. The adhesiveless case (Menga et al., 2014, 2016b, 2018a) is also reported for reference (dashed lines). In agreement with previous studies (Carbone and Mangialardi, 2004), at very low or very high sliding velocity the system recovers the elastic limit, with the high velocity solution exhibiting a smaller contact area because of the larger stiffness of the material. The most interesting result in Figs. 3(a) and (b) is that, regardless of the controlled parameter ($\tilde{\Delta}$ or $\tilde{p}_\infty$), the contact area presents a maximum located in the intermediate range of sliding velocity $\zeta$. A similar behavior has never been predicted for adhesiveless viscoelastic contacts, whereas it has been observed in several experimental tests with viscoelastic adhesive contacts as, for instance, in the case of rolling contacts between rigid cylinders and rubbery-like substrates under tractive loads (Barquins et al., 1978; Charmet and Barquins, 1996). The reported enhanced adhesion is ascribable to local viscoelastic losses occurring close to the trailing edge (crack opening) of the contact, where the excitation frequency can be qualitatively estimated as $v/\rho$, with $\rho$ being the radius of curvature at the crack tip. The analysis of the crack opening profile easily shows that $\rho \ll \lambda$ for the cases of interest (i.e., $\Gamma \ll 1$); therefore, at sliding speed $v \approx \rho/\tau \ll \lambda/\tau$, the bulk of the material behaves elastically with the low frequency elastic modulus $E_0$ (notice, the bulk excitation frequency is $\approx v/\lambda \ll 1/\tau$) and most of the energy dissipation occurs close to the trailing edge. This regime, which has been experimentally observed, is usually refereed to as the small-scale viscoelasticity regime, and also is related to friction dissipations commonly known as adhesive friction (see Section 4.1).

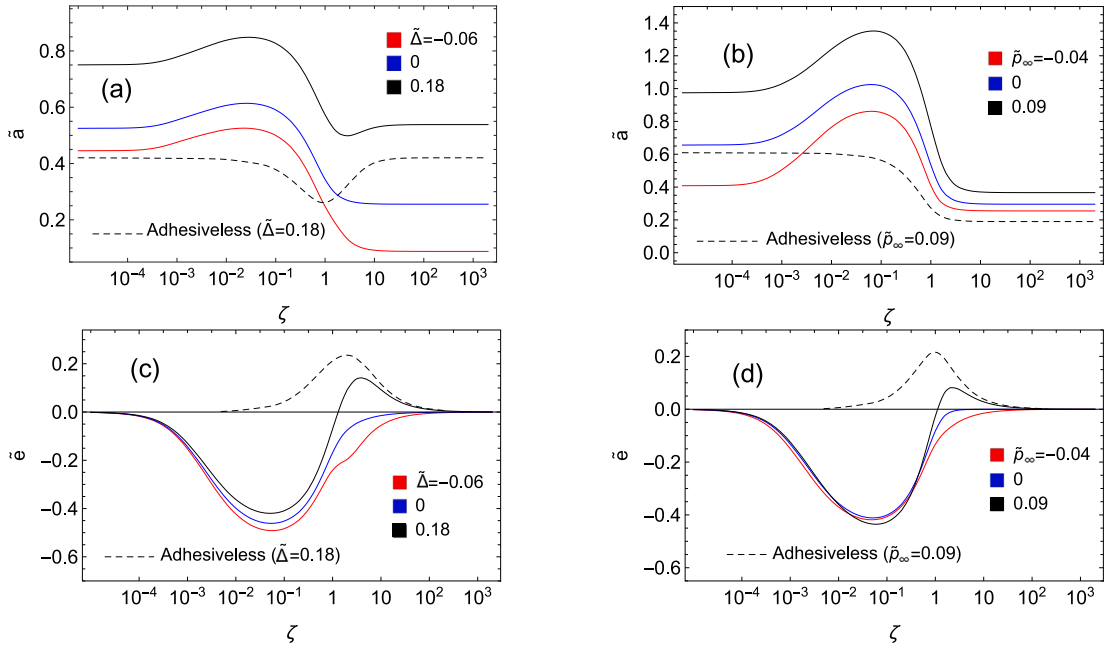

**Fig. 3.** The dimensionless semi-width of the contact $\bar{a}$, and the dimensionless eccentricity $\bar{e}$ as functions of the dimensionless sliding velocity $\zeta$, for different values of the dimensionless remote pressure $\bar{p}_\infty$, and the dimensionless penetration $\bar{\Delta}$. Results are shown for $\beta = 10$ and $\Gamma = 0.008$.

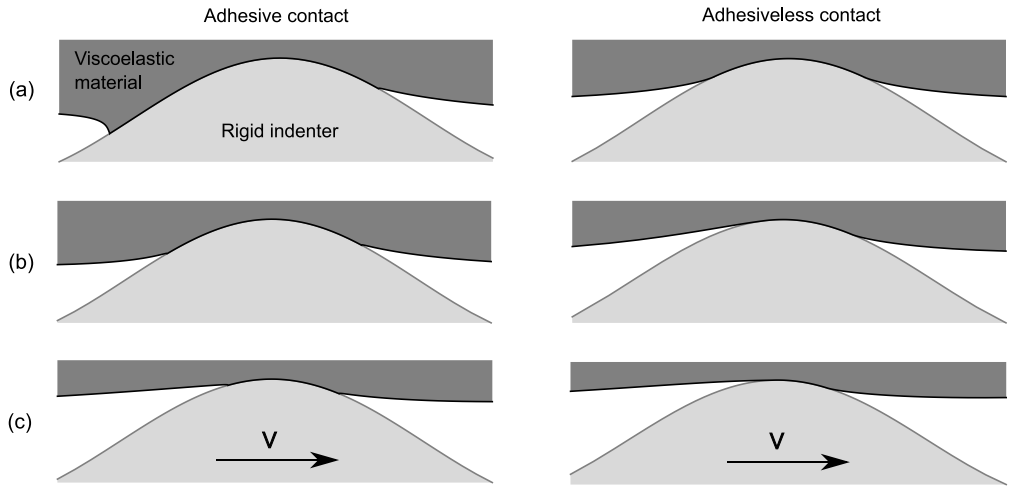

**Fig. 4.** The comparison between the deformed contact configurations predicted for adhesive and adhesiveless conditions at different sliding velocity values. (a) small-scale viscoelastic regime, $\zeta = 0.01$. (b) coupled large and small scale viscoelastic regime, $\zeta = 0.8$. (c) bulk viscoelastic regime, $\zeta = 2$. Results are shown for $\bar{p}_\infty = 0.15$, $\Gamma = 0.008$, $\beta = 10$.

In Figs. 3(c) and (d) we observe that at dimensionless speed at which the contact size $\bar{a}$ takes its maximum value, the eccentricity $\bar{e}$ is negative so that the whole contact is shifted backward. Since in the adhesiveless case the opposite behavior occurs (Menga et al., 2014, 2016b, 2018a), we conclude that in the range of velocity governed by the small-scale viscoelastic the contact area is strongly enlarged at the trailing edge. Indeed, this is confirmed by the deformed contact configuration reported in Fig. 4(a) which, in agreement with experimental observations (Barquins et al., 1978; Charmet and Barquins, 1996), suggests that in those contact conditions $v \approx \rho/\tau \ll \lambda/\tau$. However, as the sliding velocity increases and $\rho/\tau < v < \lambda/\tau$, a large amount of energy is dissipated in the bulk of the material and large-scale (bulk) viscoelastic losses take place in addition to local hysteresis at the trailing and closing edges, as qualitatively shown in Fig. 4(b)). In this case, the contact area and the eccentricity gradually invert their trend (see Figs. 3). At higher velocity, where $v \approx \lambda/\tau \gg \rho/\tau$ [see Fig. 4(c)], the contact edges behave elastically (with high frequency elastic modulus $E_\infty$), and the bulk viscoelasticity governs the system behavior.

Fig. 5 shows the effect of the dimensionless parameter $\Gamma = \bar{\gamma}/\bar{\Lambda}^2$ on the semi-width of the contact $\bar{a}$ and eccentricity $\bar{e}$. As expected, increasing $\Gamma$ enhances the effect of the small-scale viscoelasticity.

Fig. 6 reports the equilibrium values of the reduced strain energy release rates $G_1/\Delta\gamma$ (at the opening crack) and $G_2/\Delta\gamma$ (at the closing crack) as a function of the dimensionless sliding velocity $\zeta$, for different values of the dimensionless remote pressure. The effect of the dimensionless parameter $\beta = E_\infty/E_0$ on the curves is also shown in Fig. 7. As expected, at very low or very high velocities, the ratios $G_1/\Delta\gamma$ and $G_2/\Delta\gamma$ approach the unit value as the material behaves elastically. In such conditions, the non-conservative contribution to the work of internal stresses vanishes. At low velocity, increasing the value of $\zeta$ leads to an increase of $G_1/\Delta\gamma$ at the opening edge, and to a decrease of $G_2/\Delta\gamma$ at the closing edge. The difference between $G_1$ and $G_2$ is usually referred to as adhesion hysteresis, and represents a major effect of small-scale viscoelasticity. Interestingly, a maximum and a minimum value of $G_1$ and $G_2$ exist, which turn out to be monotonic functions of $\beta$, with $G_{1\,\text{max}}/\Delta\gamma < \beta$ and $G_{2\,\text{min}}/\Delta\gamma > \beta^{-1}$ (see also Fig. 7). Note that for negative value of the remote pressure $\bar{p}_\infty$ the trend of $G_1/\Delta\gamma$ and $G_2/\Delta\gamma$ is described by bell-shaped curves, with $G_1$ being always greater than $\Delta\gamma$ and $G_2$ always smaller. However, when the remote pressure $\bar{p}_\infty$ is positive, the shape of the reduced energy release rates changes. In this case $G_1$ may reach a minimum value less than $\Delta\gamma$ and $G_2$ a maximum value greater than $\Delta\gamma$. This happens because the viscoelastic stiffening of the bulk material tends to move the contact forward: a larger amount of elastic energy is stored at the closing edge compared to the opening edge, thus making $G_2$ greater than $G_1$ at sufficiently large velocity.

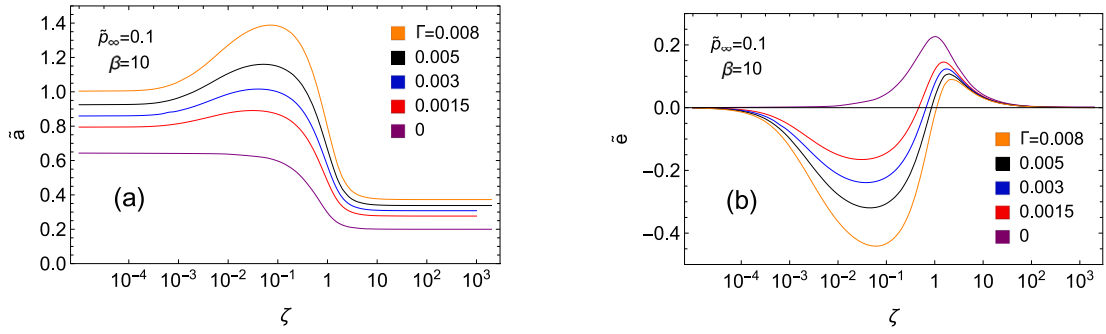

**Fig. 5.** The effect of the reduced energy of adhesion $\Gamma = \bar{\gamma}/\bar{\lambda}^2$ on the dimensionless semi-width $\bar{a}$ (a) and the dimensionless eccentricity $\bar{e}$ (b) shown as functions of the dimensionless sliding velocity $\zeta$.

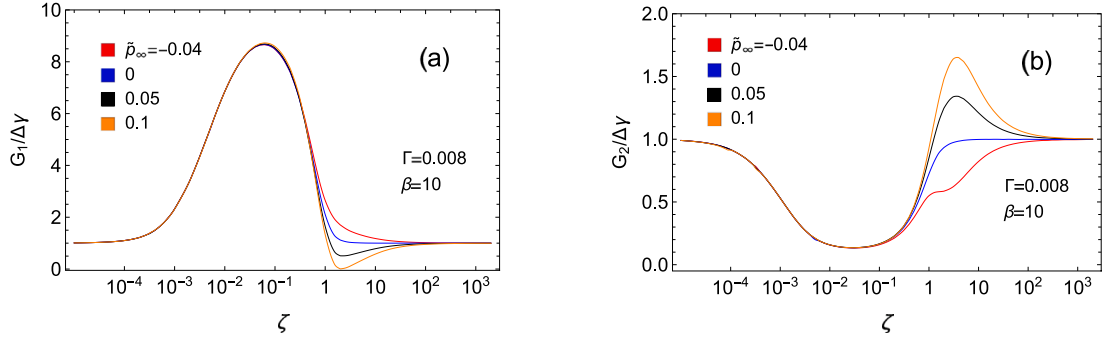

**Fig. 6.** The reduced energy release rates $G/\Delta\gamma$ as functions of the dimensionless sliding velocity $\zeta$, for different values of the dimensionless remote pressure $\bar{p}_\infty$. (a) $G_1/\Delta\gamma$ refers to the trailing edge of the contact (opening crack), (b) $G_2/\Delta\gamma$ refers to the leading edge (closing crack).

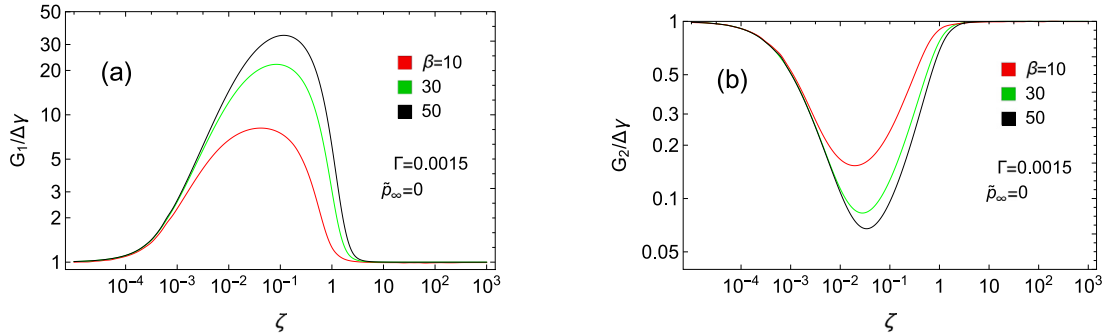

**Fig. 7.** The effect of the parameter $\beta = E_\infty/E_0$ on the trend of $G/\Delta\gamma$ vs. $\zeta$. Data are shown in log–log form. (a) $G_1/\Delta\gamma$ refers to the trailing edge of the contact (opening crack), (b) $G_2/\Delta\gamma$ refers to the leading edge (closing crack).

According to existing theories (Persson, 2021a, 2017), in the presence of a finite contact length, energy dissipation occurs also in the bulk of the material and is accompanied by material stiffening. On the contrary, in infinite systems, energy is dissipated only close to the crack tip, where the material is fully relaxed with elastic modulus equal to $E_0$. In such conditions, we may say that the system response is always governed by the small-scale viscoelasticity regime. In this case, at sufficiently high crack propagation velocity, close to the crack tip the material behaves elastically with stiffness $E_\infty$, whereas far from the crack into the bulk the material has stiffness $E_0$. Thus, one obtains $G_1/\Delta\gamma = \beta = E_\infty/E_0$ and $G_2/\Delta\gamma = \beta^{-1} = E_0/E_\infty$ (Persson and Brener, 2005; Carbone and Persson, 2005a; Greenwood, 2004; Persson, 2021b).

Notably, in Fig. 8 we show that, by increasing the size of our system (i.e. the wavelength $\lambda$), the response asymptotically approaches the one predicted for the infinite case, as $G_{1\,\text{max}}/\Delta\gamma$ and $G_{2\,\text{min}}/\Delta\gamma$ approach the values of $\beta$ and $\beta^{-1}$, respectively.

### 4.1. Friction

In this section, we investigate the frictional behavior of the system. Friction is originated by the cyclic deformations caused by the relative motion between the indenter and the solid and, in turn, by energy dissipation within the viscoelastic material. As a consequence, the contact pressure distribution $p(x)$ is asymmetric, and the resulting contact force acting on the rigid asperity leads to a tangential force opposing the indenter sliding motion, which is usually referred to as viscoelastic friction. According to Refs. Carbone and Putignano (2013), Menga et al. (2021) and Persson (2001), the viscoelastic friction coefficient can be calculated through Eq. (19). Note that dimensionless arguments yield $\mu \propto \bar{A}$, so that we can define the reduced friction coefficient as

$$\bar{\mu} = \frac{\mu}{\bar{A}} \qquad (22)$$

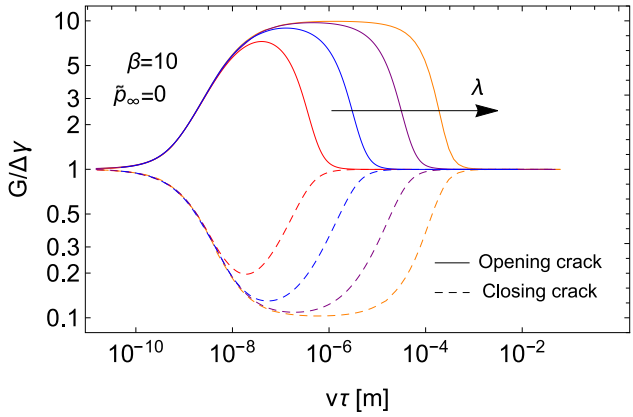

**Fig. 8.** The effect of the wavelength of the sinusoidal indenter $\lambda$ on the trend of $G/\Delta\gamma$ vs. the dimensional quantity $v\tau = \zeta/k$. Data are shown in log-log form. The wavelength increases from 10 μm (red curve) to 1.2 mm (orange curve), corresponding to $\Gamma$ increasing from 0.0001 (red curve) to 0.012 (orange curve).

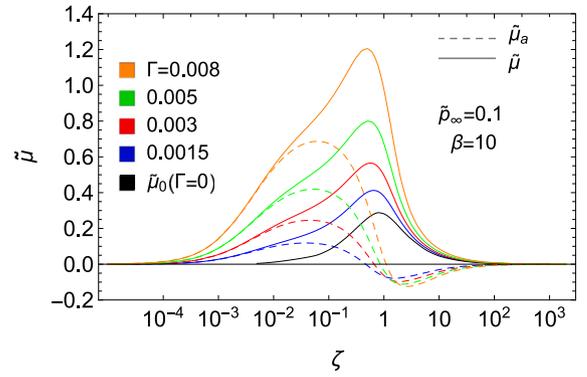

**Fig. 9.** The reduced viscoelastic friction coefficient $\bar{\mu} = \mu/\bar{A}$ (solid lines) as a function of the dimensionless sliding velocity $\zeta$, for different values of the reduced energy of adhesion $\Gamma$ under load controlled conditions. In the same figure, we also show the reduced adhesive friction coefficient $\bar{\mu}_a$ (dashed line), and the reduced friction coefficient $\bar{\mu}_0$ corresponding to adhesiveless conditions (solid black line).

The reduced friction coefficient $\bar{\mu}$ takes into account the energy dissipation occurring in the whole viscoelastic solid, i.e. both large- and small-scale viscoelastic hysteresis. In order to provide a rough estimate of the contribution to the overall friction ascribable to adhesion hysteresis (i.e., to local viscoelastic effects close to the contact edges), we define the reduced adhesive friction coefficient as

$$\bar{\mu}_a = \frac{1}{\bar{A}} \frac{G_1 - G_2}{\lambda p_\infty} \qquad (23)$$

Similarly, we also refer to $\bar{\mu}_0$ as to the reduced friction coefficient calculated in adhesiveless viscoelastic contacts, which originates from bulk viscoelastic hysteresis. Fig. 9 reports, at given remote pressure $\bar{p}_\infty$, the reduced friction coefficients $\bar{\mu}$, $\bar{\mu}_a$ and $\bar{\mu}_0$ as functions of the dimensionless sliding velocity $\zeta$, for different values of the reduced energy of adhesion $\Gamma$. At low velocity (i.e., for $\zeta < 10^{-2}$), the system is in the small-scale viscoelasticity regime; indeed, friction is governed by the adhesion hysteresis (i.e., $\bar{\mu} \approx \bar{\mu}_a$), and increasing $\Gamma$ leads to significantly higher values of $\bar{\mu}$, as the term $G_1 - G_2$ in Eq. (23) increases. At intermediate velocities (i.e., for $10^{-2} < \zeta < 1$), also bulk dissipation occurs; however, since $\bar{\mu} \approx \bar{\mu}_a + \bar{\mu}_0$, the contributions to friction of the small-scale and of the bulk (i.e. large-scale) viscoelasticity cannot be linearly separated. This is a key result: adhesion increases the contact area, hence, the volume where viscoelastic losses take place, and, in turn, increases the bulk viscoelastic dissipation [see also Fig. 3(a)]. At higher velocity (i.e., for $1 < \zeta < 10$), this effect is even clearer, as $\bar{\mu}_a < 0$ while $\bar{\mu} > \bar{\mu}_0 > 0$. Indeed, following Eq. (23), $\bar{\mu}_a$ is a qualitative estimation of the sole contribution to friction ascribable to the contact edges, which can also become negative when more energy is recovered during sliding in closing the leading edge compared to that required to open the trailing one (i.e., $G_2 > G_1$). At very high velocity, the contact edges behave almost elastically (glassy region), thus the small-scale viscoelastic energy dissipation vanishes, and the great majority of energy dissipation occurs in the bulk of the material.

Figs. 11(a, b) show the reduced friction coefficient $\bar{\mu}$ versus the sliding dimensionless speed $\zeta$ for different values of the dimensionless remote pressure $\bar{p}_\infty$ and the dimensionless penetration $\bar{\Delta}$, respectively. The specific dependence of the friction coefficient on $\bar{\Delta}$ and $\bar{p}_\infty$ is affected by different mechanisms, related to both the adhesion hysteresis and the bulk viscoelasticity. At relatively high velocity (i.e., for $\zeta > 1$),

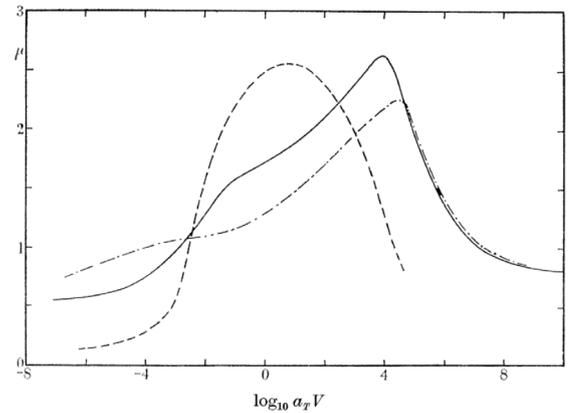

**Fig. 10.** The measured friction coefficient $\mu$ as a function of the dimensionless sliding velocity taken from Grosch (1963) for styrene–butadiene rubber sliding against three surfaces: smooth clean (dashed); rough clean (solid); rough dusted (dot-dashed). See Grosch (1963) for further details.

friction is mostly governed by bulk dissipation, and the curve $\mu$ vs. $\zeta$ roughly depends on the size of the contact area $a$, as discussed in Refs. Menga et al. (2016b, 2018a, 2021). The first effect is that, since the excitation frequency in the bulk material can be estimated as $\omega \approx 2\pi v/a = 2\pi\zeta/(\bar{a}\tau)$, and the viscoelastic dissipation takes its maximum at $\omega \approx 1/\tau$, the dimensionless sliding velocity $\zeta_0$ associated to the $\mu$ peak depends on the value of $a$ (as $\zeta_0 \approx \bar{a}/2\pi$). Indeed, Figs. 11(a, b) show that increasing $\bar{\Delta}$ or $\bar{p}_\infty$ (i.e., increasing $a$) shifts the friction peak location $\zeta_0$ at higher values. Secondly, considering that dimensional arguments yield $\mu \approx (\delta/2a)\operatorname{Im}[E(\omega)]/|E(\omega)|$, with $\delta = \Lambda(1 - \cos ka) \approx \Lambda(ka)^2/2$ being the local indenter penetration, Figs. 11(a, b) lead to higher peak values for $\mu$. On the contrary, at low velocities (i.e., for $\zeta < 10^{-2}$), most of the contribution to friction arises from adhesion hysteresis, and $\mu \approx \mu_a$. In agreement with Refs. Persson (2000) and Carbone and Mangialardi (2004), the reduction of $\mu$ reported under these condition as $\bar{\Delta}$ and $\bar{p}_\infty$ are increased, can be explained recalling that, in Eq. (23), the term $G_1 - G_2$ only depends on $\zeta$ (see also Fig. 6). Moreover, it is worth noticing, that under load controlled conditions [Fig. 9 and Fig. 11(a)], the $\mu$ vs. $\zeta$ curves present a hump localized at the value of $\zeta$ corresponding to the maximum of $\mu_a$, followed by a peak at higher velocity, where maximum bulk dissipation occurs.

In Fig. 10, we report the experimental measurements provided by Grosch in Grosch (1963) for sliding friction of rubber samples. Regardless of the numerical values, which depend on the specific rubber property and surface roughness parameters, Grosch's trends are in very good agreement with our numerical predictions. Notably, using a clean smooth surface (dashed line) as sliding counterpart, only adhesive friction occurs. Dealing with a clean rough surface (continuum line), both adhesive hysteresis and bulk viscoelasticity play a key role on $\mu$; whereas, adhesion can be completely masked by introducing a fine powder at the interface (dot-dashed line).

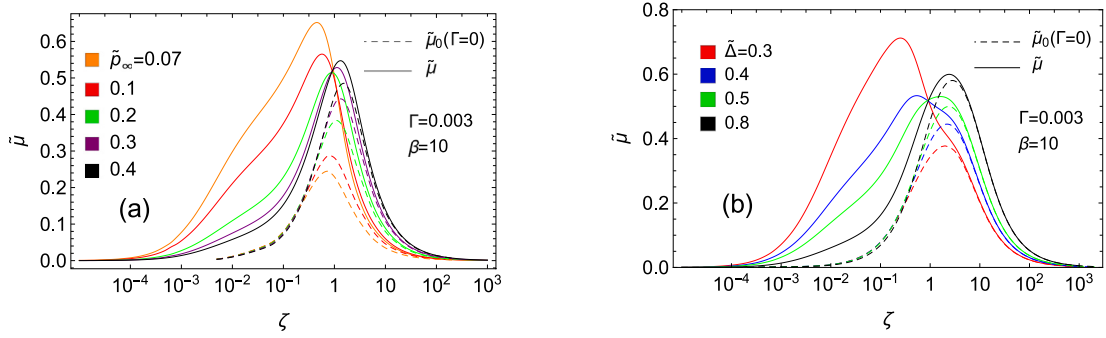

**Fig. 11.** The reduced viscoelastic friction coefficient $\tilde{\mu} = \mu/\tilde{\Lambda}$ (solid lines) as function of the dimensionless sliding velocity $\zeta$, under load controlled conditions (a) and displacement controlled conditions (b). In the same figure, we also show the reduced friction coefficient $\tilde{\mu}_0$ corresponding to adhesiveless conditions (dashed line).

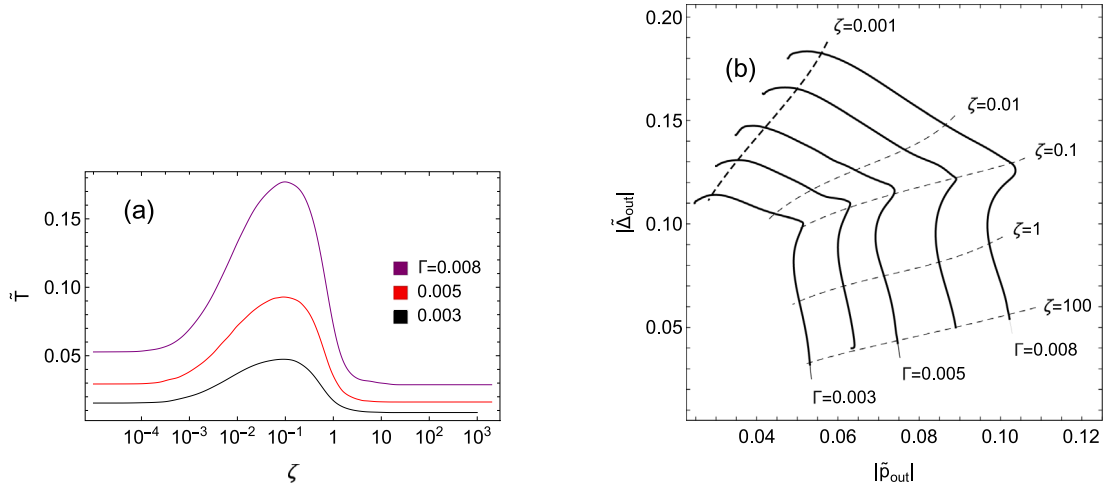

**Fig. 12.** (a): The dimensionless work needed to cause detachment $\tilde{T}$ as a function of the dimensionless sliding velocity $\zeta$, for different values of the reduced energy of adhesion $\Gamma$. (b): iso-$\zeta$ (dashed lines) and iso-$\Gamma$ (solid lines) curves in the $|\tilde{\Delta}_{out}|$ vs $|\tilde{p}_{out}|$ plane, being $\tilde{p}_{out}$ the dimensionless pull-off remote pressure and $\tilde{\Delta}_{out}$ the dimensionless penetration at which the pull-off occurs. Results are shown for $\beta = 10$.

Under displacement controlled conditions (i.e., fixed $\Delta$), the behavior is slightly different as the contact size is less affected by the effective bulk stiffness and, in turn, by the sliding velocity [see also Fig. 3(a)]. In this case, at high velocity (i.e., for $\zeta > 1$), the effect of adhesion is very poor, and $\tilde{\mu} \approx \tilde{\mu}_0$. Nonetheless, at low velocity, adhesion plays a key role as $\mu \approx \mu_a$.

### 4.2. Adhesive properties of the contact

In this section the adhesive properties of the contact are investigated in terms of toughness $T$ (i.e. amount of work required to separate the contacting bodies) and adhesive strength (i.e. pull-off remote pressure $p_{out}$) of the contact interface. In dimensionless terms the toughness $\tilde{T} = 2(1 - \nu^2) T / (\Lambda^2 E_0)$ is defined as

$$\tilde{T} = 2\pi \int_{\tilde{\Delta}_0}^{\tilde{\Delta}_{out}} \tilde{p}_{out}(\tilde{\Delta}) d\tilde{\Delta} \tag{24}$$

where $\tilde{\Delta}_{out}$ is the dimensionless penetration at which pull-off occurs, $\tilde{\Delta}_0$ is the dimensionless penetration corresponding to $\tilde{p}_{out} = 0$. Notably, both $\tilde{\Delta}_{out}$ and $\tilde{\Delta}_0$ depend on the dimensionless sliding velocity $\zeta$. In Fig. 12(a) the quantity $\tilde{T}$ is plotted against $\zeta$ for different values of the dimensionless adhesive parameter $\Gamma$, whereas in Fig. 12(b) the iso-$\zeta$ and iso-$\Gamma$ curves are shown in the $|\tilde{p}_{out}|$ vs. $|\tilde{\Delta}_{out}|$ plane. In agreement with the experimental observations reported in Ref. Charmet and Barquins (1996), we note that in the range of velocity where small-scale viscoelasticity effects take the largest values (i.e., $10^{-2} < \zeta < 10^{-1}$), both the adhesive toughness and the adhesive strength take their maximum values. Interestingly, the trend of $|\tilde{p}_{out}|$ vs. $\zeta$ curve is non-monotonic; indeed, at very high velocity, due to the glassy stiff behavior of the material, the contact interface is able to withstand high tensile loads (i.e. large pull-off pressures), with low adhesive toughness $\tilde{T}$. Interestingly, a similar limiting behavior is reported in the case of thin elastic adhesives, where the material confinement induced by the rigid substrate leads very high contact stiffness (Carbone and Mangialardi, 2008; Menga et al., 2016a).

In Fig. 13 the equilibrium diagram $\tilde{p}_\infty$ vs. $\tilde{\Delta}$ is shown, for different values of $\zeta$. We observe that $\tilde{p}_\infty$ non-monotonically depends on $\zeta$. Indeed, at low velocities (i.e., for $\zeta < 0.1 = 0.001, \zeta = 0.01, \zeta = 0.1$) the *snap into full-contact* pressure decreases as $\zeta$ is increased, as a consequence of the viscoelasticity-induced enhanced adhesion. The scenario is reversed at high velocities, as larger values of $\zeta$ entail a strong viscoelastic material stiffening.

## 5. Conclusion

In this study, we present a novel theory of adhesive viscoelastic contact mechanics in the presence of relative sliding or rolling motion between the viscoelastic solid and rigid rough indenter. While in adhesiveless conditions, the system response only depends on the bulk (large-scale) viscoelasticity with excitation frequency related to the contact size, we found that in the adhesive case the contact behavior is also affected by the local viscoelastic response close to the contact edges (i.e., small-scale viscoelasticity), where the excitation frequency depends on the adhesive (opening and closing) crack tip radii. Consequently, the overall adhesive response can be governed by either the small-, the large-scale viscoelasticity, or a combination of the two, depending on the specific value of the sliding velocity. Indeed, in agreement with the experiments (Barquins et al., 1978; Charmet and Barquins, 1996), we show that at relatively low sliding velocity the bulk of the material behaves as a soft elastic body, and the interaction between interfacial adhesion and small-scale viscoelasticity leads to an increase of the contact area, mostly localized at the trailing edge of the contact, and to a strong enhancement of the pull-off load. Small-scale viscoelasticity induces different adhesive response of the trailing and leading edge, whose difference is mostly responsible for the overall frictional response of the contact. At intermediate velocities, bulk viscoelasticity and local viscoelasticity coexist, leading to a strong increase of friction compared to the corresponding adhesiveless contact case. This peculiar result is in perfect agreement with the observations made by Grosch on rubber adhesive friction (Grosch, 1963).

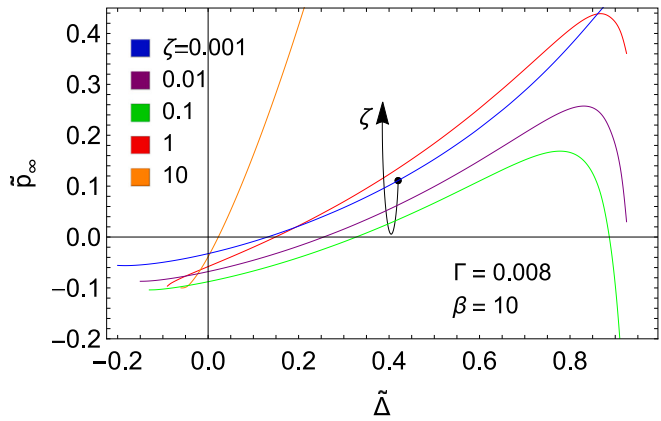

**Fig. 13.** The dimensionless remote pressure $\bar{p}_\infty$ as a function of the dimensionless penetration $\tilde{\Delta}$ at equilibrium, for different values of the dimensionless sliding velocity $\zeta$.

The present theory also allows to quantify the energy release rates $G_1$ and $G_2$ (at the trailing and leading edge, respectively) as functions of the sliding velocity. A detailed analysis of these trends shows that, because of the finiteness of the contact area, $G_1$ and $G_2$ follow a non-monotonic trend, which may also differ from the simple bell-shaped curve depending on load conditions and on the relative interplay between small-scale and bulk viscoelasticity. Interestingly, conditions exist able to completely mask adhesion at the opening edge.

**CRediT authorship contribution statement**

**C. Mandriota:** Conceptualization, Formal analysis, Investigation, Methodology, Software, Validation, Visualization, Writing – original draft, Writing – review & editing. **N. Menga:** Conceptualization, Formal analysis, Investigation, Methodology, Software, Validation, Visualization, Writing – original draft, Writing – review & editing. **G. Carbone:** Conceptualization, Formal analysis, Investigation, Methodology, Software, Validation, Visualization, Writing – original draft, Writing – review & editing.

**Declaration of competing interest**

The authors declare the following financial interests/personal relationships which may be considered as potential competing interests:

C. Mandriota reports financial support was provided by Government of Italy Ministry of Education University and Research. N. Menga reports financial support was provided by Government of Italy Ministry of Education University and Research. G. Carbone reports financial support was provided by Government of Italy Ministry of Education University and Research.

**Data availability**

Data will be made available on request.

**Acknowledgments**

This work was partly supported by the Italian Ministry of University and Research under the Programme "Department of Excellence "(decree 232/2016) and partly by the European Union - NextGenerationEU through the Italian Ministry of University and Research under the programs: (GC) National Sustainable Mobility Center CN00000023 (decree nr. 1033 - 17/06/2022), Spoke 11 – Innovative Materials Lightweighting; (NM) PRIN2022 (Projects of Relevant National Interest) grant nr. 2022SJ8HTC - ELectroactive gripper For mIcro-object maNipulation (ELFIN); (NM) PRIN2022 PNRR (Projects of Relevant National Interest) grant nr. P2022MAZHX - TRibological modellIng for sustainaBle design Of induStrial friCtiOnal inteRfacEs (TRIBOSCORE). The opinions expressed are those of the authors only and should not be considered as representative of the European Union or the European Commission's official position. Neither the European Union nor the European commission can be held responsible for them.

**Appendix A. The green function of periodic and non-periodic contacts.**

First let us recall that under the condition of translational invariance (i.e. homogeneity) and linearity, the relation between the interfacial normal stresses $\sigma(\mathbf{x}, t)$ and the surface normal displacements $u(\mathbf{x}, t)$ can be written as a convolution product, i.e.

$$u(\mathbf{x}, t) = \int d^2 x_1 dt_1 G\left(\mathbf{x} - \mathbf{x}_1, t - t_1\right) \sigma\left(\mathbf{x}_1, t_1\right) \quad \text{(A.1)}$$

where $\mathbf{x}$ is the in-plane position vector. Taking the time and space Fourier transform of Eq. (A.1) one obtains

$$u(\mathbf{q}, \omega) = G(\mathbf{q}, \omega) \sigma(\mathbf{q}, \omega) \quad \text{(A.2)}$$

where the wave vector is $\mathbf{q}$ and $\omega$ is the angular frequency, and $G(\mathbf{q}, \omega)$ is the response function. The specific form of the response function $G(\mathbf{q}, \omega)$ depends on the system geometry, on the material properties, and on how the system is constrained. Dimensional arguments (see Ref. Carbone et al., 2009a) show that $G(\mathbf{q}, \omega)$ must have the following general form

$$G(\mathbf{q}, \omega) = -\frac{2\left(1 - v^2\right)}{E(\omega)} \frac{1}{|\mathbf{q}|} S(|\mathbf{q}|, \omega) \quad \text{(A.3)}$$

where the term $S(\mathbf{q}, \omega)$ is a corrective factor that in the case of homogeneous half space is equal to 1. The corrective factor $S(\mathbf{q}, \omega)$ has been found for different geometries and different boundary conditions as well as for and also for layered materials (Ref. Carbone and Putignano, 2013; Carbone et al., 2009a; Carbone and Mangialardi, 2008; Menga et al., 2016b). In the case of a thick slab of thickness $d$ sandwiched between a flat rigid plate (upper part) and a rough substrate (bottom part), as shown in Fig. 14(a) the quantity $S(|\mathbf{q}|, \omega)$ is $\omega$-independent and takes the form

$$S(|\mathbf{q}|, \omega) = \frac{(3 - 4v)\sinh\left(2|\mathbf{q}| d\right) - 2|\mathbf{q}| d}{(3 - 4v)\cosh\left(2|\mathbf{q}| d\right) + 2\left(|\mathbf{q}| d\right)^2 - 4v(3 - 2v) + 5} \quad \text{(A.4)}$$

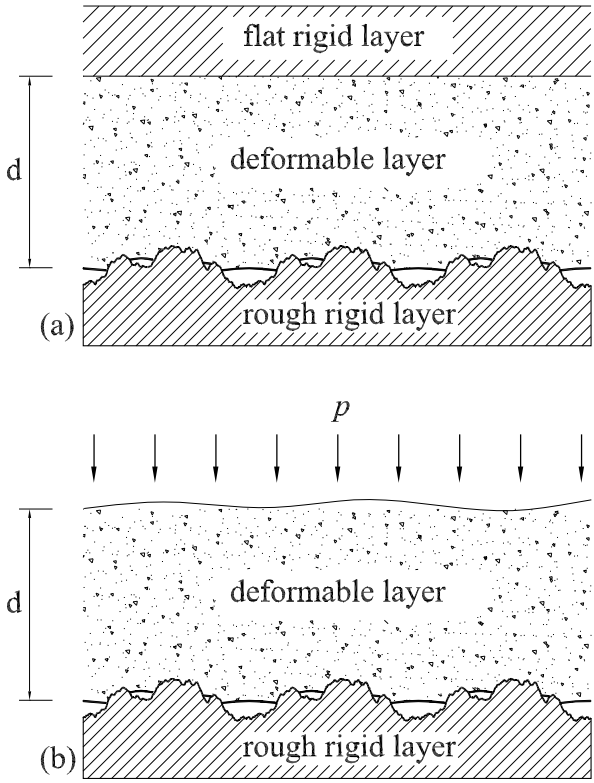

**Fig. 14.** A deformable layer of thickness $d$ in contact with a rough substrate. The layer is assumed to be glued to the upper plate (a), or subjected to a uniform pressure $p$ (b).

If we instead consider the situation depicted in Fig. 14(b), where the thick slab is subjected to a uniform applied pressure the quantity $S(|\mathbf{q}|, \omega)$ is $\omega$-independent and takes the form

$$S(|\mathbf{q}|, \omega) = \frac{\sinh(2|\mathbf{q}|d) + 2|\mathbf{q}|d}{\cosh(2|\mathbf{q}|d) - 2(|\mathbf{q}|d)^2 - 1} \tag{A.5}$$

Now let us consider steady sliding contacts. In this case, using the replacement $\mathbf{x} \rightarrow \mathbf{x} + \mathbf{v}t$, i.e. changing the reference frame so that the observer moves with velocity $\mathbf{v}$, the explicit time dependence will disappears and Eq. (A.1) can be rephrased as

$$u(\mathbf{x}) = \int d^2 x' \mathcal{G}(\mathbf{x} - \mathbf{x}', \mathbf{v}) \sigma(\mathbf{x}') \tag{A.6}$$

where the new Green function $\mathcal{G}(\mathbf{x}, \mathbf{v})$ parametrically depends on the velocity $\mathbf{v}$. Also observe that in steady sliding any physical quantities $f$ depends on space and time through the relation $f(\mathbf{x}, t) = f(\mathbf{x} - \mathbf{v}t)$. Hence, taking the Fourier transform yields

$$f(\mathbf{q}, \omega) = \int dt d^2 x \, f(\mathbf{x} - \mathbf{v}t) \, e^{-i(\mathbf{q} \cdot \mathbf{x} - \omega t)} = 2\pi \delta(\omega - \mathbf{q} \cdot \mathbf{v}) f(\mathbf{q}) \tag{A.7}$$

Then, using Eq. (A.2), and integrating over the frequency real axis gives

$$u(\mathbf{q}) = G(\mathbf{q}, \mathbf{q} \cdot \mathbf{v}) \sigma(\mathbf{q}) \tag{A.8}$$

Taking the inverse Fourier transform, Eq. (A.8) shows that

$$\mathcal{G}(\mathbf{x}, \mathbf{v}) = \frac{1}{(2\pi)^2} \int d^2 q \, G(\mathbf{q}, \mathbf{q} \cdot \mathbf{v}) \, e^{i\mathbf{q} \cdot \mathbf{x}} \tag{A.9}$$

We will show now how it is possible, moving from $\mathcal{G}(\mathbf{q}, \mathbf{v}) = G(\mathbf{q}, \mathbf{q} \cdot \mathbf{v})$ to calculate the Green function for the case of periodic steady sliding contacts, where the relation between the stress and displacement fields at the interface can be written as

$$u(\mathbf{x}, t) = \int_D d^2 x' \mathcal{G}_D(\mathbf{x} - \mathbf{x}', \mathbf{v}) \sigma(\mathbf{x}', t') \tag{A.10}$$

where $\mathcal{G}_D(\mathbf{x}, \mathbf{v})$ is the periodic Green function with periodic square cell $D$ of lateral size $L$. Of course $\mathcal{G}_D(\mathbf{x}, \mathbf{v})$ is the interfacial displacement field resulting from a stress distribution of concentrated unit loads distributed on a regular square lattice of elementary cell $D$. This distribution of forces can be represented by the surface stress field

$$\delta_D(\mathbf{x}) = \sum_{k,h=-\infty}^{+\infty} \delta\left(\mathbf{x} - \frac{2\pi}{q_0}\mathbf{k}\right) \tag{A.11}$$

where $\delta(\mathbf{x})$ is the two-dimensional Dirac delta function and $\mathbf{k} = (k, h)$ is the vectorial wave number. The fundamental frequency is $q_0 = 2\pi/L$. Therefore we get

$$\mathcal{G}_D(\mathbf{x}, \mathbf{v}) = \int d^2 x' \mathcal{G}(\mathbf{x} - \mathbf{x}', \mathbf{v}) \delta_D(\mathbf{x}') = \sum_{k,h=-\infty}^{+\infty} \mathcal{G}\left(\mathbf{x} - \frac{2\pi}{q_0}\mathbf{k}, \mathbf{v}\right) \tag{A.12}$$

Taking the Fourier transform of Eq. (A.12) gives

$$\begin{aligned} \mathcal{G}_D(\mathbf{q}, \mathbf{v}) &= \sum_{k,h=-\infty}^{+\infty} \int d^2 x e^{-i\mathbf{q} \cdot \mathbf{x}} \mathcal{G}\left(\mathbf{x} - \frac{2\pi}{q_0}\mathbf{k}, \mathbf{v}\right) \\ &= \sum_{r,s=-\infty}^{+\infty} \mathcal{G}(\mathbf{q}, \mathbf{v}) \delta\left(\frac{\mathbf{q}}{q_0} - \mathbf{r}\right) \end{aligned} \tag{A.13}$$

with $\mathbf{r} = (r, s)$. Moving back to the space domain we have

$$\mathcal{G}_D(\mathbf{x}, \mathbf{v}) = \frac{1}{(2\pi)^2} \int d^2 q \mathcal{G}_D(\mathbf{q}, \mathbf{v}) \, e^{i\mathbf{q} \cdot \mathbf{x}} = \left(\frac{q_0}{2\pi}\right)^2 \sum_{r,s=-\infty}^{+\infty} \mathcal{G}(q_0 \mathbf{r}, \mathbf{v}) \, e^{iq_0 \mathbf{r} \cdot \mathbf{x}} \tag{A.14}$$

The procedure just presented so far can easily exploited also for 1D+1D contact problems with spatial periodicity $L$ and fundamental spatial frequency $q_0 = 2\pi/L$. In this case the displacement and stress fields takes the form $u(\mathbf{x}) = u(x)$ and $\sigma(\mathbf{x}) = \sigma(x)$. The corresponding 2D Fourier transform is

$$u(\mathbf{q}) = \int dx^2 u(x) \, e^{-i\mathbf{q} \cdot \mathbf{x}} = 2\pi \delta(q_y) u(q_x) \tag{A.15}$$

and

$$\sigma(\mathbf{q}) = \int dx^2 \sigma(x) \, e^{-i\mathbf{q} \cdot \mathbf{x}} = 2\pi \delta(q_y) \sigma(q_x) \tag{A.16}$$

Therefore, after integrating over $q_y$, Eq. (A.8) gives

$$u(q_x) = \mathcal{G}^{1D}(q_x, v) \sigma(q_x) \tag{A.17}$$

where $\mathcal{G}^{1D}(q_x, v) = G(q_x, q_y = 0, q_x v)$, where $v = v_x$. Then, following the same approach as in Appendix A leads to

$$\mathcal{G}_L^{1D}(x, v) = \sum_{k=-\infty}^{+\infty} \frac{q_0}{2\pi} \mathcal{G}^{1D}(kq_0, v) \, e^{ikq_0 x} \tag{A.18}$$

## Appendix B. Non-symmetry of the Green function and the geometric interpretation of the non conservative work term $L_P$

In this Section we provide some additional considerations which emphasizes how the conservative nature of linear systems is related to the symmetry properties of the response matrix $K_{ij}$. As a simple example we consider the one represented in Fig. 15 and let us consider the application of the two forces at the points 1 and 2. Following the path I, firstly the force applied at the point 1 is slowly increased (i.e., through a quasi-static transformation) from zero to a final value $F_1$, whereas the force at the point 2 is held equal to zero. Because of linearity, the work done by the force from the configuration (a) to (b) is $L_{ab} = \frac{1}{2} u_{11} F_1$, being $u_{11}$ the displacement of the point 1 in the direction of $F_1$ in the state (b). Then, moving from (b) to (d), the force applied at the point 2 is slowly increased from zero to a final value $F_2$, whereas the force $F_1$ is held constant. At the state (d) the value of the displacements of the points 1 and 2, in the direction of $F_1$ and $F_2$, are respectively $u_1$ and $u_2$. Within the latter process, the work done by $F_2$ can be calculated by relying on linearity, whereas the work done by the constant force $F_1$ is simply obtained by multiplying $F_1$ by the relative displacement of point 1 during the process, so that the overall work from (b) to (d) is $L_{bd} = \frac{1}{2}(u_2 - u_{21})F_2 + (u_1 - u_{11})F_1$, being $u_{21}$ the displacement of the point 2 in the configuration (b). Regarding the path II, the forces are applied with reverse order. At the intermediate state (c), in which only $F_2$ is applied, the displacement of the point 1 is $u_{12}$, whereas the displacement of the point 2 is $u_{22}$. Observing that linearity entails $u_1 = u_{11} + u_{12}$ and $u_2 = u_{21} + u_{22}$ we can compare the overall work within the two paths:

$$L_I = L_{ab} + L_{bd} = \frac{1}{2}u_{11}F_1 + \frac{1}{2}u_{22}F_2 + u_{12}F_1 \tag{B.1}$$

$$L_{II} = L_{ac} + L_{cd} = \frac{1}{2}u_{22}F_2 + \frac{1}{2}u_{11}F_1 + u_{21}F_2$$

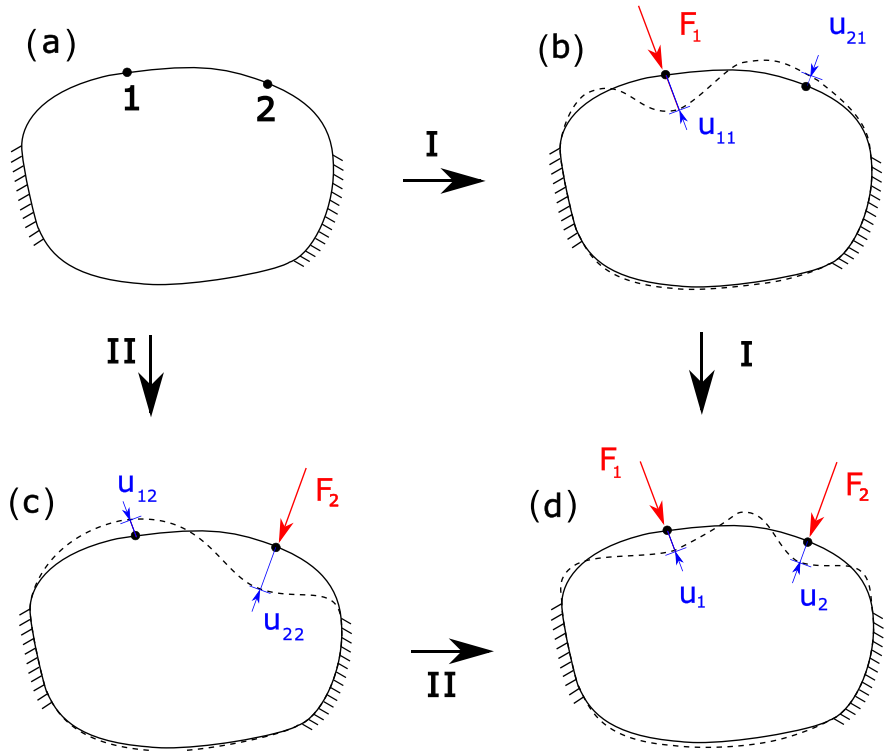

**Fig. 15.** Application of two forces at the free boundary of a constrained linear solid with reverse order. Red arrows and blue arrows refer, respectively, to forces and displacements. The work done by the forces is path-independent only when the system response matrix is symmetrical.

The system is conservative only if $L_I = L_{II}$, i.e. $u_{12}F_1 = u_{21}F_2$. This only occurs if the system's response matrix is symmetrical and can be easily shown expressing $F_1$ and $F_2$ in terms of the displacements in configurations (b) and (c) respectively and in terms of $K_{ij}$:

$$u_{12}F_1 = K_{11}u_{12}u_{11} + K_{12}u_{12}u_{21} \tag{B.2}$$

$$u_{21}F_2 = K_{21}u_{21}u_{12} + K_{22}u_{21}u_{22}$$

Moreover, observing the configuration (c), we have $F_1 = K_{11}u_{12} + K_{12}u_{22} = 0$ and similarly in configuration (b) $F_2 = K_{22}u_{21} + K_{21}u_{11} = 0$. Solving for the quantities $K_{11}u_{12} = -K_{12}u_{22}$ and $K_{22}u_{21} = -K_{21}u_{11}$ and replacing in Eq. (B.2) we conclude that the system behaves conservatively if and only if $K_{12} = K_{21}$.

Referring to the same example shown in Fig. 1, we consider now a quasi static change of the two displacements over a generic path $\mathcal{L}$ between two states, 0 and 1 in the $(u_1, u_2)$ plane [Fig. 16(a)]. We aim now at finding a geometrical interpretation for the non-conservative work $L_P$. Let us use polar coordinates: $u_1 = r(\theta)\cos\theta$ and $u_2 = r(\theta)\sin\theta$. Observe that, according to Eq. (4) and using $K_{12}^O = -K_{21}^O$, the elementary non-conservative work is $\delta L_P = K_{21}^O(u_1\delta u_2 - u_2\delta u_1)$. Recalling that $u_1\delta u_2 - u_2\delta u_1 = r^2\delta\theta$, the non-conservative work over

the whole path is:

$$L_P = \int_{\mathcal{L}} \delta L_P = K_{21}^O \int_{\theta_0}^{\theta_1} r^2\delta\theta = 2K_{21}^O A_{\mathcal{L}} \tag{B.3}$$

Where $A_{\mathcal{L}}$ is the area of the sector limited by the curve $\mathcal{L}$ and the two straight lines $\theta = \theta_0$ and $\theta = \theta_1$ [Fig. 16(a)]. If we now consider the inverse process over the same path, the work $L_P$ inverts its sign. Also note that when $\mathcal{L}$ lies on a straight line through the origin of the plane the non-conservative term is zero. Moreover, considering cyclic processes, if $K_{ij}$ is non-symmetrical the work done by the two forces over the cycle equates the non-conservative work and is thus proportional to the area $A_{\mathcal{L}_C}$ of the cycle [Fig. 16(b)] comprised by the closed curve $\mathcal{L}_C$. If the cycle is inverted, the work changes sign, leading to the conclusion that the non-conservative system might also represent a source of energy for the observer that applies the forces.

Moving to the sliding contact between a rigid rough indenter and a linear viscoelastic slab, using Eq. (1) we have

$$\delta L = \int d^2x\sigma(\mathbf{x})\delta u(\mathbf{x}) \tag{B.4}$$

$$= \int d^2x d^2x_1\mathcal{G}(\mathbf{x} - \mathbf{x}_1, \mathbf{v})\sigma(\mathbf{x})\delta\sigma(\mathbf{x}_1).$$

where $\delta\sigma(\mathbf{x})$ is an infinitesimal change of the normal stress distribution. Splitting the Green function into its odd and even parts we also get that the elastic energy can be calculated as

$$U = \frac{1}{2}\int d^2x d^2x_1\mathcal{G}^E(\mathbf{x} - \mathbf{x}_1, \mathbf{v})\sigma(\mathbf{x})\sigma(\mathbf{x}_1) \tag{B.5}$$

and the interfacial non-conservative contribution as

$$\delta L_P = \int d^2x d^2x_1\mathcal{G}^O(\mathbf{x} - \mathbf{x}_1, \mathbf{v})\sigma(\mathbf{x})\delta\sigma(\mathbf{x}_1) \tag{B.6}$$

$$= \frac{1}{2}\int d^2x d^2x_1\mathcal{G}^O(\mathbf{x} - \mathbf{x}_1, \mathbf{v})\left[\sigma(\mathbf{x})\delta\sigma(\mathbf{x}_1) - \delta\sigma(\mathbf{x})\sigma(\mathbf{x}_1)\right]$$

which vanishes in the case of purely elastic material [i.e., $\mathcal{G}^O(\mathbf{x}, \mathbf{v}) = 0$].

With reference to Eq. (B.6), we can identify some particular cases yielding $\delta L_P = 0$. We consider a perturbation of the displacements fields $u(\mathbf{x})$, which keeps its shape unchanged, i.e. $u(\mathbf{x}) = u_0(\mathbf{x})\eta$ and $\delta u(\mathbf{x}) = u_0(\mathbf{x})\delta\eta$, where $\eta$ is a dimensionless parameter governing the process. In this case, linearity yields $\sigma(\mathbf{x}) = \sigma_0(\mathbf{x})\eta$ and $\delta\sigma(\mathbf{x}) = \sigma_0(\mathbf{x})\delta\eta$, where $u_0(\mathbf{x})$ and $\sigma_0(\mathbf{x})$ are related through Eq. (1). Thus, referring to Eq. (B.6), we have $\sigma(\mathbf{x})\delta\sigma(\mathbf{x}_1) = \sigma_0(\mathbf{x})\sigma_0(\mathbf{x}_1)\eta\delta\eta = \delta\sigma(\mathbf{x})\sigma(\mathbf{x}_1)$ leading to $\delta L_{\mathrm{P}} = 0$. This arguments applies of course also to the case of a concentrated load $\sigma(\mathbf{x}) = \eta\delta_D(\mathbf{x} - \mathbf{x}_0)$, being $\delta_D(\mathbf{x})$ the Dirac delta function. Indeed, we get $\sigma(\mathbf{x})\delta\sigma(\mathbf{x}_1) = \delta_D(\mathbf{x} - \mathbf{x}_0)\delta_D(\mathbf{x}_1 - \mathbf{x}_0)\eta\delta\eta = \delta\sigma(\mathbf{x})\sigma(\mathbf{x}_1)$ leading again to $\delta L_{\mathrm{P}} = 0$.

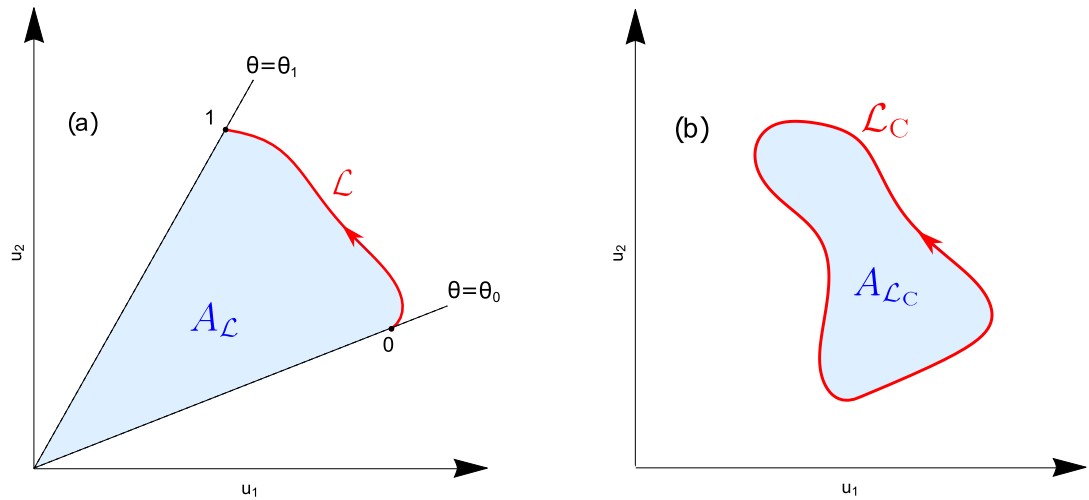

**Fig. 16.** (a) In the $(u_1, u_2)$ plane the non-conservative contribution to the work is proportional to the area of the blue sector $A_{\mathcal{L}}$. (b) In cyclic processes the overall work is proportional to the area $A_{\mathcal{L}_C}$ comprised by the closed curve and might be both positive or negative, depending on the direction the cycle is followed.